\documentclass[11pt, a4paper]{article}

\usepackage{jheppub}

\newcommand{\gf}{\mathfrak{g}}
\newcommand{\gfh}{\widehat{\gf}}

\newcommand{\del}{\partial}
\newcommand{\delb}{{\bar\partial}}

\newcommand{\ket}[1]{|#1\rangle}
\newcommand{\bra}[1]{\langle #1|}

\newcommand{\SU}{\mathrm{SU}}
\newcommand{\SO}{\mathrm{SO}}
\newcommand{\Spin}{\mathrm{Spin}}

\newcommand{\slf}{\mathfrak{sl}}
\newcommand{\suf}{\mathfrak{su}}

\newcommand{\U}{\mathrm{U}}

\newcommand{\iso}{\cong}

\newcommand{\Z}{\mathbb{Z}}

\newcommand{\R}{\mathbb{R}}
\newcommand{\C}{\mathbb{C}}

\newcommand{\T}{\mathbb{T}}
\renewcommand{\L}{\mathbb{L}}

\let\nc\newcommand
\let\renc\renewcommand
\nc{\wbar}{\overline}
\let\td\tilde
\let\wtd\widetilde
\let\wht\widehat
\let\mcl\mathcal

\nc{\ab}{{\bar{a}}} \nc{\at}{\tilde{a}} \nc{\ah}{\hat{a}}
\nc{\bb}{{\bar{b}}} \nc{\bt}{\tilde{b}} \nc{\bh}{\hat{b}}
\nc{\cb}{{\bar{c}}} \nc{\ct}{\tilde{c}} 
\nc{\db}{{\bar{d}}} \nc{\dt}{\tilde{d}} \renc{\dh}{\hat{d}}
\nc{\eb}{{\bar{e}}} \nc{\et}{\tilde{e}} \nc{\eh}{\hat{e}}
\nc{\fb}{{\bar{f}}} \nc{\ft}{\tilde{f}} \nc{\fh}{\hat{f}}
\nc{\gb}{{\bar{g}}} \nc{\gt}{\tilde{g}} \nc{\gh}{\hat{g}}
\nc{\hb}{{\bar{h}}} \nc{\hh}{\hat{h}} 
\nc{\ib}{{\bar{\imath}}} \nc{\ih}{\hat{\imath}} 
\nc{\jb}{{\bar{\jmath}}} \nc{\jt}{\tilde{\jmath}} \nc{\jh}{\hat{\jmath}}
\nc{\kb}{{\bar{k}}} \nc{\kt}{\tilde{k}} \nc{\kh}{\hat{k}}
\nc{\lb}{{\bar{l}}} \nc{\lt}{\tilde{l}} \nc{\lh}{\hat{l}}
\nc{\mb}{{\bar{m}}} \nc{\mt}{\tilde{m}} \nc{\mh}{\hat{m}}
\nc{\nb}{{\bar{n}}} \nc{\nt}{\tilde{n}} \nc{\nh}{\hat{n}}
\nc{\ob}{{\bar{o}}} \nc{\ot}{\tilde{o}} \nc{\oh}{\hat{o}}
\nc{\pb}{{\bar{p}}} \nc{\pt}{\tilde{p}} \nc{\ph}{\hat{p}}
\nc{\qb}{{\bar{q}}} \nc{\qt}{\tilde{q}} \nc{\qh}{\hat{q}}
\nc{\rb}{{\bar{r}}} \nc{\rt}{\tilde{r}} \nc{\rh}{\hat{r}}
\renc{\sb}{{\bar{s}}} \nc{\st}{\tilde{s}} \nc{\sh}{\hat{s}}
\nc{\tb}{{\bar{t}}} \renc{\th}{\hat{t}} 
\nc{\ub}{{\bar{u}}} \nc{\ut}{\tilde{u}} \nc{\uh}{\hat{u}}
\nc{\vb}{{\bar{v}}} \nc{\vt}{\tilde{v}} \nc{\vh}{\hat{v}}
\nc{\wb}{{\bar{w}}} \nc{\wt}{\tilde{w}} \nc{\wh}{\hat{w}}
\nc{\xb}{{\bar{x}}} \nc{\xt}{\tilde{x}} \nc{\xh}{\hat{x}}
\nc{\yb}{{\bar{y}}} \nc{\yt}{\tilde{y}} \nc{\yh}{\hat{y}}
\nc{\zb}{{\bar{z}}} \nc{\zt}{\tilde{z}} \nc{\zh}{\hat{z}}

\nc{\Ab}{\wbar{A}} \nc{\At}{\wtd{A}} \nc{\Ah}{\wht{A}}
\nc{\Bb}{\wbar{B}} \nc{\Bt}{\wtd{B}} \nc{\Bh}{\wht{B}}
\nc{\Cb}{\wbar{C}} \nc{\Ct}{\wtd{C}} \nc{\Ch}{\wht{C}}
\nc{\Db}{\wbar{D}} \nc{\Dt}{\wtd{D}} \nc{\Dh}{\wht{D}}
\nc{\Eb}{\wbar{E}} \nc{\Et}{\wtd{E}} \nc{\Eh}{\wht{E}}
\nc{\Fb}{\wbar{F}} \nc{\Ft}{\wtd{F}} \nc{\Fh}{\wht{F}}
\nc{\Gb}{\wbar{G}} \nc{\Gt}{\wtd{G}} \nc{\Gh}{\wht{G}}
\nc{\Hb}{\wbar{H}} \nc{\Ht}{\wtd{H}} \nc{\Hh}{\wht{H}}
\nc{\Ib}{\wbar{I}} \nc{\It}{\wtd{I}} \nc{\Ih}{\wht{I}}
\nc{\Jb}{\wbar{J}} \nc{\Jt}{\wtd{J}} \nc{\Jh}{\wht{J}}
\nc{\Kb}{\wbar{K}} \nc{\Kt}{\wtd{K}} \nc{\Kh}{\wht{K}}
\nc{\Lb}{\wbar{L}} \nc{\Lt}{\wtd{L}} \nc{\Lh}{\wht{L}}
\nc{\Mb}{\wbar{M}} \nc{\Mt}{\wtd{M}} \nc{\Mh}{\wht{M}}
\nc{\Nb}{\wbar{N}} \nc{\Nt}{\wtd{N}} \nc{\Nh}{\wht{N}}
\nc{\Ob}{\wbar{O}} \nc{\Ot}{\wtd{O}} \nc{\Oh}{\wht{O}}
\nc{\Pb}{\wbar{P}} \nc{\Pt}{\wtd{P}} \nc{\Ph}{\wht{P}}
\nc{\Qb}{\wbar{Q}} \nc{\Qt}{\wtd{Q}} \nc{\Qh}{\wht{Q}}
\nc{\Rb}{\wbar{R}} \nc{\Rt}{\wtd{R}} \nc{\Rh}{\wht{R}}
\nc{\Sb}{\wbar{S}} \nc{\St}{\wtd{S}} \nc{\Sh}{\wht{S}}
\nc{\Tb}{\wbar{T}} \nc{\Tt}{\wtd{T}} \nc{\Th}{\wht{T}}
\nc{\Ub}{\wbar{U}} \nc{\Ut}{\wtd{U}} \nc{\Uh}{\wht{U}}
\nc{\Vb}{\wbar{V}} \nc{\Vt}{\wtd{V}} \nc{\Vh}{\wht{V}}
\nc{\Wb}{\wbar{W}} \nc{\Wt}{\wtd{W}} \nc{\Wh}{\wht{W}}
\nc{\Xb}{\wbar{X}} \nc{\Xt}{\wtd{X}} \nc{\Xh}{\wht{X}}
\nc{\Yb}{\wbar{Y}} \nc{\Yt}{\wtd{Y}} \nc{\Yh}{\wht{Y}}
\nc{\Zb}{\wbar{Z}} \nc{\Zt}{\wtd{Z}} \nc{\Zh}{\wht{Z}}

\nc{\CA}{\mcl{A}} \nc{\CAb}{\wbar{\CA}} \nc{\CAt}{\wtd{\CA}} \nc{\CAh}{\wht{\CA}}
\nc{\CB}{\mcl{B}} \nc{\CBb}{\wbar{\CB}} \nc{\CBt}{\wtd{\CB}} \nc{\CBh}{\wht{\CB}}
\nc{\CC}{\mcl{C}} \nc{\CCb}{\wbar{\CC}} \nc{\CCt}{\wtd{\CC}} \nc{\CCh}{\wht{\CC}}
\nc{\CD}{\mcl{D}} \nc{\CDb}{\wbar{\CD}} \nc{\CDt}{\wtd{\CD}} \nc{\CDh}{\wht{\CD}}
\nc{\CE}{\mcl{E}} \nc{\CEb}{\wbar{\CE}} \nc{\CEt}{\wtd{\CE}} \nc{\CEh}{\wht{\CE}}
\nc{\CF}{\mcl{F}} \nc{\CFb}{\wbar{\CF}} \nc{\CFt}{\wtd{\CF}} \nc{\CFh}{\wht{\CF}}
\nc{\CG}{\mcl{G}} \nc{\CGb}{\wbar{\CG}} \nc{\CGt}{\wtd{\CG}} \nc{\CGh}{\wht{\CG}}
\nc{\CH}{\mcl{H}} \nc{\CHb}{\wbar{\CH}} \nc{\CHt}{\wtd{\CH}} \nc{\CHh}{\wht{\CH}}
\nc{\CI}{\mcl{I}} \nc{\CIb}{\wbar{\CI}} \nc{\CIt}{\wtd{\CI}} \nc{\CIh}{\wht{\CI}}
\nc{\CJ}{\mcl{J}} \nc{\CJb}{\wbar{\CJ}} \nc{\CJt}{\wtd{\CJ}} \nc{\CJh}{\wht{\CJ}}
\nc{\CK}{\mcl{K}} \nc{\CKb}{\wbar{\CK}} \nc{\CKt}{\wtd{\CK}} \nc{\CKh}{\wht{\CK}}
\nc{\CL}{\mcl{L}} \nc{\CLb}{\wbar{\CL}} \nc{\CLt}{\wtd{\CL}} \nc{\CLh}{\wht{\CL}}
\nc{\CM}{\mcl{M}} \nc{\CMb}{\wbar{\CM}} \nc{\CMt}{\wtd{\CM}} \nc{\CMh}{\wht{\CM}}
\nc{\CN}{\mcl{N}} \nc{\CNb}{\wbar{\CN}} \nc{\CNt}{\wtd{\CN}} \nc{\CNh}{\wht{\CN}}
\nc{\CO}{\mcl{O}} \nc{\COb}{\wbar{\CO}} \nc{\COt}{\wtd{\CO}} \nc{\COh}{\wht{\CO}}
\nc{\CQ}{\mcl{Q}} \nc{\CQb}{\wbar{\CQ}} \nc{\CQt}{\wtd{\CQ}} \nc{\CQh}{\wht{\CQ}}
\nc{\CR}{\mcl{R}} \nc{\CRb}{\wbar{\CR}} \nc{\CRt}{\wtd{\CR}} \nc{\CRh}{\wht{\CR}}
\nc{\CS}{\mcl{S}} \nc{\CSb}{\wbar{\CS}} \nc{\CSt}{\wtd{\CS}} \nc{\CSh}{\wht{\CS}}
\nc{\CT}{\mcl{T}} \nc{\CTb}{\wbar{\CT}} \nc{\CTt}{\wtd{\CT}} \nc{\CTh}{\wht{\CT}}
\nc{\CU}{\mcl{U}} \nc{\CUb}{\wbar{\CU}} \nc{\CUt}{\wtd{\CU}} \nc{\CUh}{\wht{\CU}}
\nc{\CV}{\mcl{V}} \nc{\CVb}{\wbar{\CV}} \nc{\CVt}{\wtd{\CV}} \nc{\CVh}{\wht{\CV}}
\nc{\CW}{\mcl{W}} \nc{\CWb}{\wbar{\CW}} \nc{\CWt}{\wtd{\CW}} \nc{\CWh}{\wht{\CW}}
\nc{\CX}{\mcl{X}} \nc{\CXb}{\wbar{\CX}} \nc{\CXt}{\wtd{\CX}} \nc{\CXh}{\wht{\CX}}
\nc{\CY}{\mcl{Y}} \nc{\CYb}{\wbar{\CY}} \nc{\CYt}{\wtd{\CY}} \nc{\CYh}{\wht{\CY}}
\nc{\CZ}{\mcl{Z}} \nc{\CZb}{\wbar{\CZ}} \nc{\CZt}{\wtd{\CZ}} \nc{\CZh}{\wht{\CZ}}

\let\eps\epsilon
\let\ups\upsilon
\let\veps\varepsilon
\let\vtht\vartheta
\let\vsgm\varsigma
\let\vphi\varphi
\let\vrho\varrho

\nc{\alphab}{\bar{\alpha}} \nc{\alphat}{\td{\alpha}} \nc{\alphah}{\hat{\alpha}}
\nc{\betab}{\bar{\beta}}   \nc{\betat}{\td{\beta}}   \nc{\betah}{\hat{\beta}} 
\nc{\gammab}{\bar{\gamma}} \nc{\gammat}{\td{\gamma}} \nc{\gammah}{\hat{\gamma}} 
\nc{\deltab}{\bar{\delta}} \nc{\deltat}{\td{\delta}} \nc{\deltah}{\hat{\delta}} 
\nc{\epsilonb}{\bar{\eps}} \nc{\epsilont}{\td{\eps}} \nc{\epsilonh}{\hat{\eps}} 
\nc{\vepsb}{\bar{\veps}}   \nc{\vepst}{\td{\veps}}   \nc{\vepsh}{\hat{\veps}} 
\nc{\zetab}{\bar{\zeta}}   \nc{\zetat}{\td{\zeta}}   \nc{\zetah}{\hat{\zeta}} 
\nc{\etab}{\bar{\eta}}     \nc{\etat}{\td{\eta}}     \nc{\etah}{\hat{\eta}} 
\nc{\thetab}{\bar{\theta}} \nc{\thetat}{\td{\theta}} \nc{\thetah}{\hat{\theta}} 
\nc{\vthetab}{\bar{\vtht}} \nc{\vthetat}{\td{\vtht}} \nc{\vthetah}{\hat{\vtht}} 
\nc{\lambdab}{\bar{\lambda}} \nc{\lambdat}{\td{\lambda}} \nc{\lambdah}{\hat{\lambda}} 
\nc{\iotab}{\bar{\iota}}   \nc{\iotat}{\td{\iota}}   \nc{\iotah}{\hat{\iota}} 
\nc{\kappab}{\bar{\kappa}} \nc{\kappat}{\td{\kappa}} \nc{\kappah}{\hat{\kappa}} 
\nc{\lmdb}{\bar{\lmd}}     \nc{\lmdt}{\td{\lmd}}     \nc{\lmdh}{\hat{\lmd}} 
\nc{\mub}{\bar{\mu}}       \nc{\mut}{\td{\mu}}       \nc{\muh}{\hat{\mu}} 
\nc{\nub}{\bar{\nu}}       \nc{\nut}{\td{\nu}}       \nc{\nuh}{\hat{\nu}} 
\nc{\xib}{\bar{\xi}}       \nc{\xit}{\td{\xi}}       \nc{\xih}{\hat{\xi}} 
\nc{\pib}{\bar{\pi}}       \nc{\pit}{\td{\pi}}       \nc{\pih}{\hat{\pi}} 
\nc{\vpib}{\bar{\vpi}}     \nc{\vpit}{\td{\vpi}}     \nc{\vpih}{\hat{\vpi}} 
\nc{\rhob}{\bar{\rho}}     \nc{\rhot}{\td{\rho}}     \nc{\rhoh}{\hat{\rho}} 
\nc{\vrhob}{\bar{\vrho}}   \nc{\vrhot}{\td{\vrho}}   \nc{\vrhoh}{\hat{\vrho}} 
\nc{\sigmab}{\bar{\sigma}} \nc{\sigmat}{\td{\sigma}} \nc{\sigmah}{\hat{\sigma}} 
\nc{\vsigmab}{\bar{\vsgm}} \nc{\vsigmat}{\td{\vsgm}} \nc{\vsigmah}{\hat{\vsgm}} 
\nc{\taub}{\bar{\tau}}     \nc{\taut}{\td{\tau}}     \nc{\tauh}{\hat{\tau}} 
\nc{\upsilonb}{\bar{\ups}} \nc{\upsilont}{\td{\ups}} \nc{\upsilonh}{\hat{\ups}} 
\nc{\phib}{\bar{\phi}}     \nc{\phit}{\td{\phi}}     \nc{\phih}{\hat{\phi}} 
\nc{\varphib}{\bar{\vphi}}   \nc{\varphit}{\td{\vphi}}   \nc{\varphih}{\hat{\vphi}} 
\nc{\chib}{\bar{\chi}}     \nc{\chit}{\td{\chi}}     \nc{\chih}{\hat{\chi}} 
\nc{\psib}{\bar{\psi}}     \nc{\psit}{\td{\psi}}     \nc{\psih}{\hat{\psi}} 
\nc{\omegab}{\bar{\omega}} \nc{\omegat}{\td{\omega}} \nc{\omegah}{\hat{\omega}} 

\nc{\Gammab}{\wbar{\Gamma}}     \nc{\Gammat}{\wtd{\Gamma}}     \nc{\Gammah}{\wht{\Gamma}}
\nc{\Deltab}{\wbar{\Delta}}     \nc{\Deltat}{\wtd{\Delta}}     \nc{\Deltah}{\wht{\Delta}}
\nc{\Thetab}{\wbar{\Theta}}     \nc{\Thetat}{\wtd{\Theta}}     \nc{\Thetah}{\wht{\Theta}}
\nc{\Lambdab}{\wbar{\Lambda}}   \nc{\Lambdat}{\wtd{\Lambda}}   \nc{\Lambdah}{\wht{\Lambda}}
\nc{\Xib}{\wbar{\Xi}}           \nc{\Xit}{\wtd{\Xi}}           \nc{\Xih}{\wht{\Xi}}
\nc{\Pib}{\wbar{\Pi}}           \nc{\Pit}{\wtd{\Pi}}           \nc{\Pih}{\wht{\Pi}}
\nc{\Sigmab}{\wbar{\Sigma}}     \nc{\Sigmat}{\wtd{\Sigma}}     \nc{\Sigmah}{\wht{\Sigma}}
\nc{\Upsilonb}{\wbar{\Upsilon}} \nc{\Upsilont}{\wtd{\Upsilon}} \nc{\Upsilonh}{\wht{\Upsilon}}
\nc{\Phib}{\wbar{\Phi}}         \nc{\Phit}{\wtd{\Phi}}         \nc{\Phih}{\wht{\Phi}}
\nc{\Psib}{\wbar{\Psi}}         \nc{\Psit}{\wtd{\Psi}}         \nc{\Psih}{\wht{\Psi}}
\nc{\Omegab}{\wbar{\Omega}}     \nc{\Omegat}{\wtd{\Omega}}     \nc{\Omegah}{\wht{\Omega}}

\renewcommand{\psit}{\tilde\psi}
\renewcommand{\psib}{\bar\psi}

\title{On the six-dimensional origin of the AGT correspondence}
\author{Junya Yagi}
\affiliation{Department of Mathematics, University of Hamburg, \\
Bundesstrasse 55, 20146 Hamburg, Germany}
\emailAdd{junya.yagi@math.uni-hamburg.de}

\abstract{We argue that the six-dimensional $(2,0)$ superconformal
  theory defined on $M \times C$, with $M$ being a four-manifold and
  $C$ a Riemann surface, can be twisted in a way that makes it
  topological on $M$ and holomorphic on $C$.  Assuming the existence
  of such a twisted theory, we show that its chiral algebra contains a
  W-algebra when $M = \R^4$, possibly in the presence of a
  codimension-two defect operator supported on $\R^2 \times C \subset
  M \times C$.  We expect this structure to survive the
  $\Omega$-deformation.}


\arxivnumber{1112.0260}

\begin{document}
\maketitle
\flushbottom

\section{Introduction}

In recent years there have been remarkable advances in our
understanding of $\CN = 2$ supersymmetric gauge theories in four
dimensions.  One highlight is a conjectural relation between these
theories and conformal field theories (CFT) in two dimensions,
formulated by Alday, Gaiotto and Tachikawa \cite{Alday:2009aq} (AGT)
and generalized thereafter by others \cite{Wyllard:2009hg,
  Mironov:2009by, Taki:2009zd, Braverman:2010ef, Wyllard:2010rp}.  In
essence, the AGT conjecture asserts a correspondence between certain
quantities in two types of theories.  One is $\CN = 2$ theory obtained
by compactifying the six-dimensional $(2,0)$ superconformal theory on
a Riemann surface $C$ \cite{Witten:1997sc, Gaiotto:2009hg,
  Gaiotto:2009we}.  The other is CFT on $C$ with W-algebra symmetry
\cite{Bouwknegt:1995ag}.

In view of the nature of the $\CN = 2$ theory involved, it is clear
that the AGT correspondence should have its origin in six dimensions.
A nice explanation would be as follows.  Take the $(2,0)$ theory
defined on the product $\R^4 \times C$.  Compactified on $C$, the
theory reduces to an $\CN = 2$ theory on $\R^4$.  If instead we
somehow ``compactify'' it on $\R^4$, we get a theory on $C$.  This
latter theory is, presumably, a CFT with W-algebra symmetry.  Unlike
an overall scaling of the metric, scaling the metric of $\R^4$ or $C$
separately is not a symmetry.  Certain quantities are, however,
protected under separate compactification, hence can be computed in
either effective theory.  The comparison would then lead to the
correspondence.

This scenario sounds plausible, and there are pieces of evidence
supporting its valid\-ity~\cite{Alday:2009qq, Bonelli:2009zp,
  Tachikawa:2011dz}.  Nevertheless, it seems that we still lack a
satisfactory explanation of how exactly the alleged W-algebra arises
from six dimensions.  In this paper we address this~issue.

Our setup is the following.  We consider the $(2,0)$ theory of type
$\gf$ on $M \times C$, where $\gf$ is a simply-laced real simple Lie
algebra and $M$ is a four-manifold.  If $M$ and $C$ are curved, to
preserve at least one supersymmetry we need to twist the theory.  In
section 2, we argue that the theory becomes topological on $M$ and
holomorphic on $C$ after the twisting is done; in other words, it
depends on the geometry of the spacetime only through the smooth
structure of $M$ and the conformal structure of $C$.  Roughly
speaking, this means that we obtain a chiral CFT on $C$ ``with values
in a topological field theory on $M$'' \cite{Moore:2011ee}.  The
chiral algebra of this theory is the object of our interest.

In section 3, we show that the chiral algebra contains a W-algebra
when $M = \R^4$, if the twisted theory indeed has the above property.
The type of the W-algebra is precisely the one relevant for the AGT
correspondence, namely the one that results from the quantum
Drinfeld-Sokolov reduction \cite{MR1071340, FigueroaO'Farrill:1990dz,
  deBoer:1993iz} of the affine Lie algebra $\gfh$ with respect to a
principal $\slf_2$ embedding; see appendix for a brief review of
quantum Drinfeld-Sokolov reduction.  Since the chiral algebra is
protected under compactification, the same W-algebra symmetry must be
present in the effective theory on $C$, justifying the crucial
assumption in the aforementioned argument.  Our reasoning also applies
to the case where there are a number of supersymmetric codimension-two
defect operators inserted at points on $C$.  This is actually part of
the original conjecture.  Furthermore, we can place another such
defect operator on $\R^2 \times C \subset M \times C$, in which case
the relevant W-algebra changes.  This covers the generalization of the
conjecture proposed in \cite{Braverman:2010ef, Wyllard:2010rp}, which
involves $\CN = 2$ theories with a surface operator extending along
$\R^2 \subset \R^4$.

Therefore, by studying the chiral algebra of the $(2,0)$ theory, we
gain a fairly clear picture of the origin of W-algebras that appear in
the AGT correspondence and its generalization incorporating surface
operators.  We leave some questions unanswered, however.  The most
important one is about the $\Omega$-deformation
\cite{Nekrasov:2002qd}.  For a complete treatment, we must tame
infrared divergences coming from the noncompactness of the spacetime.
We will assume that this is done by some regularization procedure,%
\footnote{Without a regularization one may still find a W-algebra in
  the classical sense.  For instance, classical W-algebras can be
  obtained by turning on the $\Omega$-deformation and then removing
  it.  In the presence of the $\Omega$-deformation, it is known that
  the level of $\gfh$ is given by $k = -h^\vee +
  \epsilon_2/\epsilon_1$, where $h^\vee$ is the dual Coxeter number of
  $\gf$ and $\epsilon_1$, $\epsilon_2$ are deformation parameters.  In
  the limit $\epsilon_1 \to 0$, we have $k \to \infty$ and the quantum
  W-algebra reduces to a classical W-algebra.  The opposite limit
  $\epsilon_2 \to 0$ is also a classical limit, via the duality of
  W-algebras sending $k + h^\vee \to (k + h^\vee)^{-1}$
  \cite{Feigin:1991wy}.  (The limit in which both $\epsilon_1$ and
  $\epsilon_2$ are taken to zero but $\epsilon_2/\epsilon_1$ remains
  finite appears to be more subtle.)  In general, the level would
  depend on a specific regularization scheme that we employ.}
but the AGT conjecture picks a particular one.  That is to turn on the
$\Omega$-deformation, which confines quantum excitations within an
effectively compact region in $\R^4$.  What special role does the
$\Omega$-deformation play in our story, other than merely providing an
infrared regulator?  Although we propose a possible explanation in
section 4, a definitive answer will have to wait until the appearance
of a six-dimensional realization of the $\Omega$-deformation.

Lastly, let us point out that in principle our chiral algebra can be
much larger than just the W-algebra, and may contain other interesting
structures.  It deserves to be explored more deeply.

\acknowledgments

I thank Meng-Chwan Tan and J\"org Teschner for illuminating
discussions, and Gregory Moore, Yuji Tachikawa, and Edward Witten for
helpful comments.  I am also grateful to the support of the Center for
Frontier Science at Chiba University and the hospitality of the Simons
Center for Geometry and Physics at Stony Brook University.  This work
is supported by Deutsche Forschungsgemeinschaft through the Research
Training Group 1670 ``Mathematics Inspired by String Theory and QFT.''

\section{Chiral algebra from the (2,0) theory}

Let us see how a chiral algebra can arise from the $(2,0)$ theory on
$M \times C$.  First of all, we need to twist the theory.

The theory has the R-symmetry $\Spin(5)_{R}$ under which the sixteen
supercharges transform in the spinor representation.  Due to the
product structure of the spacetime the holonomy is $\Spin(4)_{M}
\times \Spin(2)_{C} \iso \SU(2)_{l} \times \SU(2)_{r} \times \U(1)_C$.
Correspondingly, we split $\Spin(5)_{R}$ into $\Spin(3)_{R} \iso
\SU(2)_{R}$ and $\Spin(2)_{R} \iso \U(1)_{R}$.  Under $\SU(2)_{l}
\times \SU(2)_{r} \times \U(1)_{C} \times \SU(2)_{R} \times
\U(1)_{R}$, the supercharges transform as
\begin{equation}
  \bigl((2,1)_{{1/2}} \oplus (1,2)_{{-1/2}}\bigr)
  \otimes \bigl(2_{{1/2}} \oplus 2_{{-1/2}}\bigr).
\end{equation}
The twisting is done in two steps.  The first step is to identify the
diagonal $\U(1)_{C}' \subset \U(1)_{C} \times \U(1)_{R}$ with the
holonomy group of $C$.  This gives eight supercharges that are scalars
on $C$ and so preserved by the curvature of $C$.  These transform
under $\SU(2)_{l} \times \SU(2)_{r} \times \SU(2)_{R}$~as
\begin{equation}
  (2,1,2) \oplus (1,2,2).
\end{equation}
Thus we get $\CN = 2$ supersymmetry on $M$ (which is generally broken
by the curvature of $M$).  The second step is to replace $\SU(2)_{r}$
by the diagonal $\SU(2)_{r}' \subset \SU(2)_{r} \times \SU(2)_{R}$.
Then we are left with one supercharge that is a singlet under
$\SU(2)_{l} \times \SU(2)_{r}' \times \U(1)_{C}'$.  We call it $Q$; it
has $\U(1)_R$ charge $1/2$.

It is crucial to understand whether $Q$ obeys $Q^2 = 0$ or not.  One
way to determine this is to note that if we compactify the theory on
$C$ right after the first step, then the second step is nothing but
the familiar Donaldson-Witten twist \cite{Witten:1988ze} applied to
the $\CN = 2$ theory on $M$.  In that case we know that $Q^2$ is not
zero, but equal to the gauge transformation generated by an adjoint
scalar $\sigma$.  For the $\U(1)_R$ charges to match, $\sigma$ must
come from a scalar $\Phi$ of charge $1$ in the $(2,0)$ theory.  After
the twisting, $\Phi$ is a one-form which can be written as $\Phi =
\Phi_z dz$, where $z$ is a local holomorphic coordinate on $C$.  Hence
we expect that in six dimensions, $Q^2$ is given by some sort of gauge
transformation specified by the one-form $\Phi$, that reduces in four
dimensions to the gauge transformation by $\sigma$.

In the abelian case, such a symmetry is indeed known.  The abelian
$(2,0)$ theory has a two-form ``gauge field'' $B$ (with values in a
vector bundle) whose field strength $H = dB$ is self-dual.  (More
precisely, $B$ is something called a connection on an abelian gerbe.)
The symmetry in question acts by
\begin{equation}
  \label{B}
  B \to B + d\Lambda
\end{equation}
with $\Lambda$ a one-form.  If $B_{iz}$ is identified with the
component $A_i$ of the gauge field $A$ of the $\CN = 2$ theory on $M$,
this transformation reduces to the ordinary gauge transformation $A
\to A + d_M\Lambda_z$ upon dimensional reduction on $C$.  Since there
are no other conceivable symmetry to which $Q^2$ can be equated, we
should have $Q^2 = 0$ up to the two-form gauge transformation
\eqref{B} with $\Lambda = \Phi$.

How to generalize the above symmetry to the nonabelian case is a bit
mysterious.  Still, given the fact that the nonabelian theory can be
perturbed to an abelian theory or compactified to a nonabelian $\CN =
2$ theory, it is very likely that a generalization does exist.  So let
us assume that we have $Q^2 = 0$ up to some ``gauge transformation.''
With this relation at hand, we can now define the $Q$-cohomology of
``gauge-invariant'' operators or states.  The $Q$-cohomology classes
of operators and states are the physical objects in the twisted
theory.

Since the twisted theory can be compactified to a topologically
twisted $\CN = 2$ theory on $M$, we expect it to be also topological
on $M$.  We can see how the physics depends on the geometry of $C$ by
reversing the steps in the twisting.  If we twist first along $M$,
then we get two supercharges that are scalars on $M$.  These transform
under $\U(1)_{C} \times \U(1)_{R}$ as
\begin{equation}
  (-1/2, \pm 1/2),
\end{equation}
showing that we have $(0,2)$ supersymmetry in two dimensions.  The
twisting along $C$ then turns one of the supercharges into a scalar on
$C$, which we call $Q$.  Twisted $(0,2)$ theories have the
antiholomorphic degrees decoupled.  Hence, we expect that the twisted
theory is holomorphic on $C$.

The theory being topological on $M$ and holomorphic on $C$,
$Q$-cohomology classes of local operators are independent of the
position in $M$ and vary holomorphically on $C$.  Moreover, two of
them can be multiplied by operator product expansion (OPE), with the
coefficients being holomorphic functions on $C$.  Therefore, these
$Q$-cohomology classes form a chiral algebra, an OPE algebra of
holomorphic fields, in the sense of two-dimensional CFT.  The locality
on $M$ actually plays no role here, so we may also include in the
chiral algebra $Q$-cohomology classes of operators that are local on
$C$ but nonlocal on $M$.  Such nonlocal operators will be important to
us.

So far we have made two assumptions (apart from the very existence of
the $(2,0)$ theory and some of their properties), that we have $Q^2 =
0$ up to some ``gauge transformation'' by $\Phi$, and that the twisted
theory is topological on $M$ and holomorphic on $C$.  A strong support
for these assumptions comes from the existence of an analogous twist
for $\CN = 2$ superconformal gauge theory on the product $\Sigma
\times C$ of two Riemann surfaces, introduced by Kapustin
\cite{Kapustin:2006hi} in the course of generalizing geometric
Langlands duality.  When $M = \Sigma \times \Sigma'$, our twist
reduces to that of Kapustin via compactification on $\Sigma'$.
Kapustin's theory has $Q^2 = 0$, which is consistent with our formula
for the abelian case since the transformation~\eqref{B} for $\Lambda =
\Phi$ gives a trivial gauge transformation when dimensionally reduced
to $\Sigma \times C$.  Furthermore, Kapustin's theory is topological
on $\Sigma$ and holomorphic on $C$.

\section{Identifying the W-algebra}

Our chiral algebra originated from twisted $(0,2)$ supersymmetry in
two dimensions.  This is encouraging, because the chiral algebra of a
twisted $(0,2)$ theory often contains a W-algebra.  For example, the
chiral algebra of the A-model contains the Virasoro algebra, which is
the $W_{2}$ algebra.  (In fact, it contains the $(2,0)$ superconformal
algebra \cite{Malikov:1998dw}.)  A more interesting example is
provided by the theory obtained from the A-model by killing the
left-moving fermions.  If we take the target space to be the flag
manifold of a simple Lie group, the chiral algebra of this model
contains the corresponding affine Lie algebra of critical level at the
level of perturbation theory \cite{Malikov:1998dw, Witten:2005px}.%
\footnote{Nonperturbatively, the chiral algebra vanishes once
  instanton corrections are taken into account \cite{Witten:2005px,
    Tan:2008mi, Yagi:2010tp}.  The chiral algebra of the flag manifold
  model apparently has an intimate connection with geometric Langlands
  duality \cite{Tan:2007ej, Tan:2008ak, Tan:2011uz}.}

That said, we ask: how does a W-algebra associated to $\gfh$ arise in
the chiral algebra of the twisted theory for $M = \R^4$?

To answer this question, let us think of the $(2,0)$ theory as if it
were a gauge theory, with gauge group $G$ whose Lie algebra is $\gf$,
and with all the fields valued in the adjoint representation.  We
choose a framing of field configurations at the infinity of $\R^4$.
That is to say, we regard two configurations to be physically
identical if and only if they are related by a gauge transformation
that is identity at infinity.  Let $\CG^\infty$ be the group of gauge
transformations that are global on $\R^4$, which is just the group of
maps from $C$ to $G$.  This is a physical symmetry of the theory,
rather than a gauge symmetry.  As we now see, the conserved currents
associated with this symmetry give rise to the affine currents of
$\gfh$ in the chiral algebra.

Let $\{t_a\}$ be an antihermitian basis of $\gf$, and consider a gauge
transformation\linebreak $\exp(\epsilon f t_a) \in \CG^\infty$ with $\epsilon$
an infinitesimal parameter and $f$ a real function on $C$.  Under this
transformation the gauge field changes by $\delta A = \epsilon f[t_a,
A] + \epsilon d_Cf t_a$.  To find the corresponding conserved current,
we promote $\epsilon$ to an arbitrary function supported on an open
set in $\R^4 \times C$, and look at the coefficient of $d\epsilon$ in the
variation of the action.  By gauge invariance, we may as well compute
the variation under the transformation $\delta A = \epsilon d_Cf t_a -
d(\epsilon f) t_a = d\epsilon ft_a$, which is the difference between
the transformation with $\epsilon$ promoted to a function afterwards
(hence no longer a symmetry) and the gauge transformation
$\exp(\epsilon ft_a)$.  This makes it clear that the conserved current
takes the form $fj_a$.  Define operators
\begin{equation}
  J_a = \int_{\R^4} \star_4 j_{a,z}, \qquad
  \Jb_a = \int_{\R^4} \star_4 j_{a,\zb},
\end{equation}
where $\star_4$ is the four-dimensional Hodge star operator, sending
$1$ to the volume form of $\R^4$.  If we choose a local cylindrical
coordinate $w = \sigma + i\tau$ such that $z = \exp(-iw)$ and regard
$\tau$ as time, then the integral
\begin{equation}
  \label{intfJ}
  \int \! d\sigma \, f (zJ_a + \zb\Jb_a)
\end{equation}
gives the conserved charge.  Since $Q$ is a gauge singlet, this
commutes with $Q$ for any choice of $f$, which implies that $z J_a +
\zb\Jb_a$ is $Q$-closed.  But $J_a$ and $\Jb_a$ are functionals of the
fields with no explicit dependence on the coordinate, so it must be
that they are both $Q$-closed.  Then $\Jb_a$ must be $Q$-exact.  For
it transforms nontrivially under antiholomorphic scaling on $C$, which
would contradict the holomorphy of the twisted theory if it did not
vanish in the $Q$-cohomology.  Going back to the expression
\eqref{intfJ} and setting $f = 1$, we find that the zero mode of $J_a$
acts by $t_a$ in the $Q$-cohomology.

Therefore, for each $t_a$, we obtain in the chiral algebra a
holomorphic current $J_a$ whose zero mode acts by $t_a$.  The
collection $\{J_a\}$ of such currents generate $\gfh$, as promised.  A
similar construction was in fact found by Johansen
\cite{Johansen:1994ud} in the context of holomorphically twisted $\CN
= 1$ theories defined on the product of two Riemann surfaces.  For
those theories, each flavor symmetry gives rise to an affine Lie
algebra in the $Q$-cohomology.

We have given a heuristic argument that the chiral algebra of the
$(2,0)$ theory on $\R^4 \times C$, framed at infinity, contains $\gfh$
as a subalgebra.  Of course, the $(2,0)$ theory is not really a gauge
theory, and the above reasoning does not apply as it is.  But it does
apply once we compactify the theory on a circle --- for then we have
five-dimensional maximally supersymmetric Yang-Mills theory with
gauge group $G$!  Hence, if $M = S^1 \times \R^3$ for example, we do
get the affine currents $J_a$, but this time defined by integration
over $\R^3$ in the five-dimensional theory.

For $M = \R^4$, however, a similar construction may not give the whole
$\gfh$.  One way to compactify the theory on a circle in $\R^4$ is to
bend $\R^4$ into the product of a cigar and $\R^2$, and make the cigar
very narrow.  Such a geometry was considered in \cite{Nekrasov:2010ka}
in relation to the quantum Hitchin system, and also in
\cite{Witten:2011zz} in relation to Khovanov homology.  This procedure
reduces $\R^4$ to $\R_+ \times \R^2$, so introduces a boundary at the
origin of the half-line $\R_+$.  The boundary condition here is not
our choice; it is specified by the six-dimensional theory since there
was no boundary at the beginning.  Global symmetries can change this
boundary condition.  This is not a problem if one wants to deal with
the conserved currents placed at a point away from the boundary, in
which case one derives Ward identities by considering local
transformations supported in the neighborhood of that point.  But if
one considers the currents placed on the boundary, the Ward identities
are no longer the same because of the boundary contribution.  This is
the situation we face if we define the $J_a$ by integration over $\R_+
\times \R^2$, a submanifold which intersects with the boundary.  In
order not to spoil our argument, we should project out those $J_a$
that act nontrivially on the boundary state created by the
compactification.  It is this projection that we will find implements
quantum Drinfeld-Sokolov reduction.

Thus, we are in need of understanding the boundary conditions of the
maximally \linebreak supersymmetric Yang-Mills theory.  There are
really two kinds of boundaries in the \linebreak spacetime $\R_+
\times \R^2 \times C$.  One is the boundary coming from six
dimensions.  This is $\{\infty\} \times \R^2 \times C \cup \R_+
  \times \{\infty\} \times C$, located at the infinity of $\R_+
\times \R^2$.  The other is the boundary created by the
compactification.  This is $\{0\} \times \R^2 \times C$, located at
the origin of $\R_+$.  Relevant to the projection is only the latter,
as the conserved currents used to define the $J_a$ are never placed at
infinity which, strictly speaking, is not part of the spacetime.  But
understanding the boundary condition for the former tells us something
important about the latter, so we turn to it first.

At infinity, we impose half-BPS boundary condition (that is,
preserving half of the supersymmetry) because we want $\CN = 2$
supersymmetry when we compactify the theory on $C$.  This is given as
follows \cite{Gaiotto:2009hg}.  The R-symmetry $\Spin(5)_R$ acts on
the five scalars $\phi_i$ of the theory by $\SO(5)$ rotations.  In
order to twist the theory, we have split $\Spin(5)_R$ into
$\Spin(3)_R$, rotating $(\phi_1, \phi_2, \phi_3)$, and $\Spin(2)_R$,
rotating $(\phi_4, \phi_5)$.  After the twisting, $\phi_4 + i\phi_5$
becomes the component of a $(1,0)$-form $\varphi$ along $C$.  The
boundary condition at infinity is that $A_i = \phi_i = 0$ for $i = 1$,
$2$, $3$, and for the remaining components the pair $(A, \varphi)$
solves the Hitchin equations on $C$:
\begin{equation}
  \label{Hitchin}
    F_A + [\varphi, \varphi^\dagger] = 0, \qquad
    \delb_A\varphi = 0.
\end{equation}
Notice that this boundary condition is invariant under
$\Spin(3)_R$.

The boundary condition at the origin of $\R_+$ is determined by the
boundary condition of the original six-dimensional theory, whose
dimensional reduction we just described.  As such, it must be
invariant under any symmetry (except the gauge symmetry, which is not
a physical symmetry) preserved by the boundary condition at infinity,
especially $\Spin(3)_R$.  It turns out that we have a slightly unusual
condition here: as the three scalars $\phi_1$, $\phi_2$, $\phi_3$
approach the origin $y = 0$, they develop a singularity according to
the Nahm equation
\begin{equation}
  \frac{d\phi_i}{dy} + \epsilon_{ijk} [\phi_j,\phi_k] = 0.
\end{equation}
More precisely, the three scalars must behave near $y = 0$ as
\begin{equation}
  \label{BC_y=0}
  \phi_i \sim \frac{t_i}{y}
\end{equation}
up to an $\Spin(3)_R$ transformation, where the $t_i \in \gf$ form a
standard basis of a principal $\suf_2$ subalgebra satisfying the
commutation relations $[t_i, t_j] = \epsilon_{ijk} t_k$
\cite{Witten:2011zz, Gaiotto:2008sa}.

Now we look for a W-algebra inside the chiral algebra.  From our
previous discussion, we guess that it is generated by those $J_a$ that
kill the boundary state at the origin of $\R_+$, specified by the Nahm
pole \eqref{BC_y=0}.  The residue of the pole defines via
complexification an embedding $\rho\colon \slf_2 \to \gf_\C$, which is
a basic ingredient of quantum Drinfeld-Sokolov reduction.  This
strongly suggests that the W-algebra associated to the pair
$(\gfh,\rho)$ is hidden somewhere in the chiral algebra.

Obvious candidates for the generators are the $J_a$ corresponding to
the elements of $\CG^\infty$ that leave the Nahm pole invariant, but
these only generate the $\slf_2$-invariant subalgebra of $\gfh$.  We
can find more generators if we exploit the $\Spin(3)_R$ freedom in the
boundary condition.  To see what this freedom makes the total set of
generators be, we use a little trick.  First, we enlarge the theory by
a subgroup $\CH$ of the complexification $\CG^\infty_\C$ of
$\CG^\infty$.  By this we mean the following.  The boundary conditions
define in the field configuration space a subspace over which the path
integral is performed.  We consider the orbit of this subspace in the
complexified field space generated by the action of $\CH$, and perform
the path integral over it.  After we enlarge the theory, we gauge
$\CH$ to account for the overcounting.  For this operation not to
change the physics, no real elements of $\CH$ should leave invariant
the boundary conditions (of the pre-enlarged theory); otherwise,
gauging $\CH$ would identify different configurations satisfying the
same boundary conditions, which are physically distinct in the
original description.  And yet, for the resulting gauge-fixed algebra
to be nicely described, we want $\CH$ to be large enough so that the
whole $\gfh$ lies in the chiral algebra of the enlarged theory.

Let us define $t_\pm = t_1 \pm it_2$ and $t_0 = t_3$ so that they obey
the standard $\slf_2$ commutation relations, and split $\gf_\C$ as
$\gf_\C = \gf_+ \oplus \gf_0 \oplus \gf_-$ according to the eigenvalue
of $t_0$, just as one does in quantum Drinfeld-Sokolov reduction.
Then, a good choice of $\CH$ is $\CG^\infty_+$, the subgroup of
$\CG^\infty_\C$ corresponding to $\gf_+$.  Clearly its elements change
the boundary condition at $y = 0$, so enlarging by $\CG^\infty_+$ and
gauging it does not change the physics.  Moreover, the enlarged theory
has the whole $\gfh$ in the chiral algebra.  To see this, note that
$J_a \in \gfh_+$ lie in the chiral algebra since the boundary
condition of the enlarged theory is manifestly invariant under
$\CG^\infty_+$, and so does $J_- = 2J_1 - J_+$ since $J_1$ acts on the
Nahm pole by an $\Spin(3)_R$ transformation rotating around the
$1$-axis.  Thus, by applying $J_-$ successively on $J_a \in \gfh_+$,
we can generate all the $J_a$ that belong to an $\slf_2$ multiplet of
nonzero highest weight.  Adding the $\slf_2$ singlets, we obtain
$\gfh$.

Now gauging $\CG^\infty_+$, we find that the chiral algebra contains a
subalgebra given by the BRST cohomology computed in $\gfh$.  We can
fix the gauge, for example, by setting the $\gf_+$-valued part of the
gauge field $A$ to zero at infinity.  The gauge that is directly
related to quantum Drinfeld-Sokolov reduction is defined by requiring
that the operator $J = J^a t_a$ take the form
\begin{equation}
  J = t_+ + \sum_{t_a \in \gf_0 \oplus \gf_-} J^a t_a.
\end{equation}
Assuming that the level $k$ of $\gfh$ is nonzero and finite (which we
believe is generically true from the AGT conjecture), $J/k$ transforms
under $\CG^\infty_+$ in the same way as $A_z$ does, so we can first
set the $\gf_+$-valued part of $J$ to zero.  Then we can set $J^+ = 1$
to reach the desired form, locally on $C$.  To set $J^+ = 1$ globally,
in general we need to twist the theory further by identifying the
diagonal of $\U(1)_C \times \U(1)_\rho$ with the holonomy of $C$,
where $\U(1)_\rho$ is the $\U(1)$ subgroup of $\CG^\infty$ generated
by $t_0$.  After that, $J^+$ can be thought of as a section of a
trivial bundle and equated with a global section.  The residual gauge
freedom is the antiholomorphic elements of $\CG^\infty_+$ which we
could apply right after we set the $\gf_+$-part of $J$ to zero.  We
can fix it if we want, but this is not necessary.

The above gauge-fixing procedure precisely reproduces%
\footnote{To recast the standard BRST formalism (see
  \cite{Polchinski:1998rq} for example) in the form used in
  \cite{deBoer:1993iz}, one sets the gauge-fixing condition $F^\alpha
  = -2i(J^\alpha - \delta^{\alpha+})$ for $t_\alpha \in \gf_+$, adds
  the BRST transformation of $f^{\alpha\beta}{}_\gamma b_\alpha
  b_\beta c^\gamma + b_\alpha B^\alpha$ to the action, and integrates
  $B_\alpha$ out.}
the constraints imposed by the quantum Drinfeld-Sokolov reduction of
$\gfh$ with respect to the principal $\slf_2$ embedding $\rho$.
Therefore, the gauge-fixed algebra is the W-algebra associated to
$(\gfh, \rho)$.  This was what we wanted to see.

Our argument readily generalizes to the case where the $(2,0)$ theory
has a number of half-BPS codimension-two defect operators
\cite{Gaiotto:2009hg, Gaiotto:2009we} inserted at points on $C$.  This
is simply because we can define the $J_a$ away from those points.  (If
we place $J_a$ in the neighborhood of one of the insertion points,
then we can use the Ward identity to deduce the action of $J_a$ on the
defect operator there.)  In the five-dimensional description, the
defects create singularities in the gauge field so that the fields
transform by nontrivial monodromies as they go around the insertion
points.  Defect operators of this type change the $\CN = 2$ theory on
the four-dimensional side of the AGT correspondence, while introducing
vertex operators on the two-dimensional side.

We can insert yet another defect operator at the tip of the cigar.
Upon compactification on $C$, this one becomes a half-BPS surface
operator in the $\CN = 2$ theory.  It creates a singularity of the
form $A \sim \alpha d\theta$ at the tip, where $\alpha$ is in the Lie
algebra of a maximal torus $\T \subset G$ and $\theta$ is the
azimuthal coordinate of the cigar.  The gauge group is broken on the
surface to the maximal subgroup $\L \subset G$ commuting with
$\alpha$, called the Levi subgroup.  If the theory is instead
compactified on the circle of the cigar, the monodromy is lost but
something else happens: the Nahm pole changes.  For example, when
there is a ``full'' surface operator \cite{Alday:2010vg} for which $\L
= \T$, the Nahm pole is zero and we get $\gfh$ in the chiral algebra,
as is consistent with the results found in \cite{Braverman:2004vv,
  Alday:2010vg}.  In general, it is believed that the Nahm pole in the
presence of a defect operator is one whose $t_+$ is a principal
nilpotent element in the complexification of the Lie algebra of $\L$.%
\footnote{I thank Yuji Tachikawa for explaining this point to me.}
(In the case of $\gf = A_N$, this conclusion was essentially obtained
in \cite{Gaiotto:2008ak}.)  This explains the appearance of the
W-algebra associated to this $\slf_2$ embedding in the generalization
of the AGT correspondence proposed in \cite{Braverman:2010ef,
  Wyllard:2010rp}.

\section{Role of the \texorpdfstring{$\mathbf\Omega$}{Omega}-deformation}

Even though we have identified the W-algebra in the chiral algebra of
the $(2,0)$ theory, one mystery remains: what is the role of the
$\Omega$-deformation in our story?  We conclude this paper by giving a
possible answer to this question.

Suppose that we can introduce some operation in the $(2,0)$ theory on
$\R^4 \times C$ that reduces to the $\Omega$-deformation when the
theory is compactified on $C$.  However that is realized, this
``six-dimensional $\Omega$-deformation'' must exploit in some way or
another the rotations in two orthogonal two-planes in $\R^4$.  Then,
it would modify the relation $Q^2 = 0$ (modulo a ``gauge
transformation'') by adding to the right-hand side the conserved
charges generating these rotations.  This is what happens in the case
of the usual $\Omega$-deformation in four dimensions.  To define the
$Q$-cohomology in such a situation, we have to project the algebra of
operators to the subalgebra of $Q^2$-closed operators.  This
projection would be harmless if the $(2,0)$ theory were a gauge
theory, so that we could define the $J_a$ directly in six dimensions
in a manner that is manifestly rotation invariant.  In reality, we
need first go down to five dimensions, whereby we lose one of the
directions in which the $\Omega$-deformation is performed.  Because of
this reduction it is far from obvious whether $J_a$ would survive the
$\Omega$-deformation or not.

There is, however, a sufficient condition for a given $J_a$ to
survive.  In going down to five dimensions, we took $M$ to be the
product of a cigar and $\R^2$, and sent the radius of the cigar to
zero.  For definiteness, suppose that the cigar was made of a
half-cylinder $\R_+ \times S^1$ capped with a hemisphere.  We could
turn on the $\Omega$-deformation on the cigar using the rotations
around its axis.  A peculiar property of the $\Omega$-deformation is
that we can cancel such a deformation on the flat cylinder part by a
change of variables \cite{Nekrasov:2010ka}.  In this ``undeformed''
description, the effect of the $\Omega$-deformation localizes on the
hemisphere, so the $\Omega$-deformed $Q$ obeys $Q^2 = V$ for some
conserved charge $V$ whose current is supported there.  After we
compactify the theory, the cigar becomes $\R_+$ and some boundary
state $\ket{\Psi}$ appears at $y = 0$.  Since $V$ is now supported at
$y = 0$, the statement that $J_a$ is $V$-closed is equivalent to
saying that $\bra{\Psi'} [V, J_a] \ket{\Psi} = 0$ for any states
$\ket{\Psi'}$ placed at a $y$-slice infinitesimally close to $y = 0$.
Noting that $\ket{\Psi}$ is $Q$- and hence $V$-closed as the boundary
conditions of the original theory were chosen that way, we see that
this happens if $J_a \ket{\Psi} = 0$.  Therefore, $J_a$ is $V$-closed
if it kills the boundary state created by the compactification.

So it seems to be a good idea to project out those $J_a$ that act
nontrivially on $\ket{\Psi}$.  But this projection is exactly what we
did to avoid the boundary contribution!  Then, this could be a more
fundamental reason as to why we should carry out the projection: to
define the chiral algebra in the presence of the $\Omega$-deformation.

In fact, in the absence of the $\Omega$-deformation or defect
operator, we could do away with the projection altogether by choosing
a different way to compactify the theory.  For example, the $\U(1)$
action on $\R^4 \iso \C^2$ defined by $(z_1, z_2) \to (e^{i\theta}
z_1, e^{i\theta} z_2)$ gives a smooth quotient without boundary,
$\C^2/\U(1) \iso \R^3$.  For such a choice the issue of emergent
boundary does not occur, and we expect to get the full $\gfh$.%
\footnote{As a variation of this construction, take $M$ to be the ALE
  space obtained by a hypek\"ahler resolution of the orbifold
  $\C^2/\Z_k$.  This space may be thought of as an $S^1 \times
  \R$-fibration over $\C$, where the fiber at $z \in \C$ is given by
  the equation $u^2 + v^2 = - \prod_{i=1}^k (z - a_i)$ with $u$, $v$,
  $a_i \in \C$ and the $a_i$ distinct.  The quotient by the
  $S^1$-action is $\R^3$.  So again, we expect to get $\gfh$.  This
  expectation fits nicely with I. Frenkel's conjecture that the
  cohomology of the moduli space of framed $G$-instantons on
  $\C^2/\Z_k$ should carry a level~$k$ representation of $\gfh$, which
  has been partially proved by Licata \cite{MR2737777} and led
  Braverman and Finkelberg \cite{MR2656088} to propose geometric
  Langlands duality for complex surfaces.  See \cite{Dijkgraaf:2007sw,
    Hollands:2009ar, Witten:2009at, Tan:2008wp} for physical
  explanations on this point.  In the case $k = 1$, we have argued
  that turning on the $\Omega$-deformation reduces $\gfh$ (which now
  has a different level) to a W-algebra.  For a general value of $k$,
  it should reduce $\gfh$, possibly combined with other subalgebras of
  the chiral algebra, to a parafermionic W-algebra
  \cite{Belavin:2011pp, Nishioka:2011jk}.}
One of the motivations behind our choice was that it would make our
construction compatible with this hypothetical $\Omega$-deformation of
the $(2,0)$ theory.

\appendix

\section{Quantum Drinfeld-Sokolov reduction}

Given an affine Lie algebra $\gfh$ and an embedding $\rho\colon \slf_2
\to \gf_\C$, quantum Drinfeld-Sokolov reduction \cite{MR1071340,
  FigueroaO'Farrill:1990dz, deBoer:1993iz} produces a vertex algebra
isomorphic, up to a shift in level, to a subalgebra of the universal
enveloping algebra $U(\gfh)$ of $\gfh$.  This is achieved by imposing
constraints on the affine currents by means of BRST gauge fixing.  The
resulting algebra is the W-algebra associated to the pair $(\gfh,
\rho)$, denoted by $W(\gfh, \rho)$.

The choice of the $\slf_2$ embedding gives a decomposition of $\gf_\C$
into $\slf_2$ multiplets.  Take a basis $\{t_+, t_0, t_-\}$ of the
$\slf_2$ subalgebra satisfying
\begin{equation}
  [t_+, t_-] = 2t_0, \qquad
  [t_0, t_\pm] = \pm t_\pm,
\end{equation}
and extend it to a complete basis $\{t_a\}$ of $\gf_\C$.  We assume
that $t_a$ have an integer spin (that is, $[t_0, t_a] = s_a t_a$ for
some $s_a \in \Z$), and write $\gf_+$, $\gf_0$, and $\gf_-$ for the
subalgebras of spin positive, zero, and negative, respectively.  Then
the constraints imposed by the quantum Drinfeld-Sokolov reduction are
\begin{equation}
  J^a = 0
\end{equation}
for all $t_a \in \gf_+$ except $t_+$, and
\begin{equation}
  J^+ = 1.
\end{equation}
Here we have raised the index of $J_a$ using the Killing form.  These
constraints may be thought of as coming from gauge fixing a certain
variant of gauged WZW model \cite{Balog:1990mu}, hence can be imposed
using the BRST formalism.  The BRST cohomology computed in $U(\gfh)$
is $W(\gfh,\rho)$.

One subtlety in the above procedure is that the constraint $J^+ = 1$
breaks conformal invariance if we use the standard Sugawara
energy-momentum tensor $T$, under which $J_a$ have conformal weight
one.  We remedy this problem by adding $-\del J_0$ to $T$.  This
shifts the conformal weight of $J^a$ by $-s_a$, thereby making $J^+$
weight zero.

\providecommand{\href}[2]{#2}\begingroup\raggedright\endgroup
\end{document}